\documentclass[preprint,11pt,review]{elsarticle}

\usepackage{amssymb}
\usepackage{graphics}
\usepackage{adjustbox}

\usepackage{lineno,hyperref}
\usepackage[table]{xcolor}
\usepackage{amsmath,amsfonts,amsthm,amssymb}
\usepackage{graphicx}
\usepackage[margin=1in]{geometry}
\usepackage[section]{placeins}
\usepackage{subcaption}
\usepackage{gensymb}
\graphicspath{{figures/}}
\usepackage[separate-uncertainty,table-align-uncertainty,multi-part-units=repeat]{siunitx}

\usepackage{xargs}
\usepackage{booktabs,caption}
\usepackage[flushleft]{threeparttable}

\usepackage[T1]{fontenc}
\usepackage{palatino}

\journal{Computer Methods and Programs in Biomedicine}

\begin{document}

\begin{frontmatter}



\title{\textbf{Exploring the Interplay of Left Coronary Tree Anatomy and Haemodynamics: Implications for Plaque Formation}}


\author[label1]{Mingzi Zhang, Ph.D.\corref{cor1}}
\ead{mingzi.zhang@unsw.edu.au}
\author[label1]{Ramtin Gharleghi, Ph.D}
\author[label1]{Chi Shen, M.S.}
\author[label1]{Susann Beier, Ph.D.}
\affiliation[label1]{organization={School of Mechanical and Manufacturing Engineering,
Faculty of Engineering, University of New South Wales},
            city={Kensington},
            postcode={NSW 2052}, 
            state={New South Wales},
            country={Australia}}\cortext[cor1]{Corresponding Author, Ainsworth Building, High St., Sydney, NSW Australia\\\hphantom~~~~~Phone No: +61 478 899 561}

\begin{abstract}

\textbf{Background and Objective}: The link between atherosclerosis and blood flow-induced haemodynamic luminal shear stresses is well established. However, this understanding has not been translated into clinical practice because of the interdependent effects of the complex coronary anatomy and a multitude of potential haemodynamic metrics, which have been challenging to delineate. Key questions remain, including clarification on the dominant surrogate markers of coronary plaque's focal onset and progression. Thus, this study aims to identify anatomical and haemodynamic differences in coronary trees at different stages of stenoses.

\textbf{Methods}: A total of 39 left coronary trees which are publicly available (ASOCA datset). Each coronary tree was dissected into bifurcations and non-bifurcating segments for comparisons. We calculated the inflow angle, bifurcation angle, and Finet's ratio for all bifurcations and the average diameter, torsion, and absolute curvature for both non- and bifurcating segments. Transient coronary flow simulations revealed the normalised luminal area exposed to Low Time-Average Endothelial Shear Stress (\%LowTAESS), High Oscillatory Shear Index (\%HighOSI), and High Relative Residence Time (\%HighRRT). We statistically investigated the differences between non-stenosed (\textit{n}=20, Diameter Stenosis DS=0\%), moderately (\textit{n}=12, 0\%\textless DS\textless70\%), and severely (\textit{n}=7, DS$\geq$70\%) stenosed cases, accounting for multiple comparisons and sample size, after which \textit{p}<0.05* is considered significant. 

\textbf{Results}: Only the average curvature and \%HighOSI differed between the non-stenosed (DS=0\%), and moderately or severely stenosed (DS\textgreater0\%) for the coronary trees (\textit{p}=0.024* and \textit{p} \textless0.001*), and non-bifurcating segments (\textit{p}=0.027* and \textit{p} \textless0.001*), with AUCs of 0.711 and 0.813 for the classification performance. However, the absolute OSI values were comparatively small and should be interpreted cautiously. \%LowTAESS@0.5Pa and \%HighRRT@2.5Pa\textsuperscript{-1} significantly differed between moderately (0\%\textless DS\textless 70\%) and severely (DS$\geq$70\%) stenosed trees (\textit{p}=0.009* and \textit{p}=0.012*). Classification performance or segment-specific analysis could not be performed due to the sample size (\textit{n}<15).

\textbf{Conclusions}: Our findings suggest curvature and potentially \%HighOSI being critical factors in coronary plaque onset in non-bifurcating segments, whereas \%LowTAESS and \%HighRRT affect plaque progression after onset. This new knowledge may inform future studies and highlight the merit of anatomical and haemodynamic surrogate markers for atherosclerotic risk stratification in coronary artery diseases.\\

\end{abstract}

\begin{graphicalabstract}

\includegraphics[width={0.85\textwidth}]{./Images/graphical_abstract}

\end{graphicalabstract}

\begin{highlights}
\item Vascular curvature and HighOSI appear to be key drivers in the onset of coronary plaque.
\item LowTAESS and highRRT may govern plaque progression after onset to severe stenosis. 
\item Bifurcation-only studies may not be able to capture this mechanism. 
\item Stenosed bifurcations differ in side branch diameter and distal main vessel torsion.
\item Common variations in haemodynamic thresholds did not appear to affect these relationships. 
\end{highlights}


\end{frontmatter}

\section{INTRODUCTION}

Coronary Artery Disease (CAD) is a leading cause of morbidity and mortality worldwide. The most common form of CAD is atherosclerotic plaque formations in the epicardial arteries, which may rupture, leading to acute coronary syndrome \cite{Bentzon2014Mechanisms}, or become severely obstructive, which limits distal blood supply and causes chronic myocardium ischaemia and death \cite{Xaplanteris2018Five-Year}. Therefore, early identification of patients prone to coronary atherosclerosis and their risk stratification may offer opportunities to mitigate the impact and burden of CAD.

It is well established that a patient’s age, sex, smoking history, high-density and general cholesterol levels, and blood pressure affect atherogenetic processes. However, there are great variations among individuals in terms of the plaque progression rate \cite{indraratna2022plaque}, and even for confirmed intermediate lesions, the best timing and interventional strategy are still to be explored \cite{benatti2023timing}. Together, nearly 25\% of cardiovascular events remain unexplained by our understanding to date \cite{damen_performance_2019}. Coronary anatomy and blood flow-induced haemodynamic measures, such as endothelial or wall shear stress (ESS), are considered important factors compounding this residual risk \cite{Morbiducci2016Atherosclerosis}. Yet, delineating which anatomical and haemodynamic metrics, or their combination, would cause plaque development or advance disease stages remains unanswered due to the multi-faceted complexity of the problem.  

The low and/or oscillatory shear theory has been the prevalent mechanism attributed to the focal initiation of atherosclerosis within the cardiovascular system \cite{Peiffer2013Does}. Thus, a multitude of blood flow metrics have been proposed, including the Time-Averaged Endothelial Shear Stress (TAESS) \cite{Hartman2021definition,dai2019Associations}, Oscillatory Shear Index (OSI) \cite{Morbiducci2016Atherosclerosis, Ku1985Pulsatile}, Relative Residence Time (RRT) \cite{Hoi2011Correlation}, and Wall Shear Stress (WSS) topological skeleton \cite{Morbiducci2020Wall, Chiastra2022Coronary}. 

Since the vascular haemodynamics are imposed by the local arterial anatomy \cite{Morbiducci2016Atherosclerosis}, various anatomical features have also been investigated for their efficacy as surrogate markers of pathophysiological phenomena, most notably in clinical literature, coronary tortuosity, best described as average curvature,  \cite{Kashyap2022Accuracy,Han2022Association}, and more recently, the coronary artery volume index \cite{Benetos2020Coronary} — a
ratio of coronary artery volume to myocardial mass, predicting the likelyhood of cardiovascular events.

Previous research has anatomically described the whole coronary tree \cite{Medrano-Gracia2016computational}. To our knowledge, there has been limited comprehensive description of the haemodynamics in whole coronary trees however, most studies focused on the bifurcations \cite{Medrano-Gracia2016computational} or stenotic main branches \cite{stone_prediction_2012} only. Moreover, only a few observations exist on arterial anatomy and haemodynamics before and after plaque onset, wherein the efficacy of WSS and its derivatives in localising plaques was demonstrated  \cite{Knight2010Choosing,Rikhtegar2012Choosing}. 

It is also important to note that accurate quantification of coronary anatomy and haemodynamics requires confidence in the image-based reconstruction. In fact, coronary segmentation uncertainty can be as high as 30\% \cite{Gharleghi2022Automated}. Here, we uniquely use a high-quality dataset \cite{Gharleghi2023Annotated}, comprised of segmentations verified by multiple experts to minimise inter-observer variability. 

The aim of this work is to compare the coronary tree anatomy and haemodynamics of symptomatic patients without significant luminal obstruction to patients with lumen-protruding plaques. Using the coronary trees available from the Automated Segmentation Of normal and diseased Coronary Arteries (ASOCA) Challenge \cite{Gharleghi2022Automated}, we studied non-stenosed, moderately, and severely stenosed cases. 
 
\section{METHODS}

\subsection{Study Population and Coronary Segmentation}
The open source ASOCA dataset \cite{Gharleghi2022Automated} is comprised of 40 CAD cases, of which we excluded one severe case due to extreme stenosis in the Left Anterior Descending (LAD) artery. Of the 39 left main coronary trees studied here, 20 had no significant stenosis, 12 were moderately stenosed, and 7 had severe stenoses.

This dataset is based on Computed Tomography Coronary Angiogram (CTCA) images obtained using a GE LightSpeed 64 slice CT Scanner with an ECG-gated retrospective acquisition protocol. The in-plane resolution of the acquired images was 0.3 to 0.4 mm, and the out-of-plane resolution was 0.625mm \cite{Medrano-Gracia2016computational}. All images were segmented by three experts independently using 3D Slicer \cite{Fedorov20123D}  before applying an automated majority method to derive the final vascular mask. Refer to Gharleghi \textit{et al.} \cite{Gharleghi2023Annotated} for a detailed description of the vascular reconstruction process and \textbf{Table 1}. \ref{tbl:demographics} (i) for the patient demographics. 

\begin{table}
\centering
\resizebox{0.9\textwidth}{!}{
    \begin{threeparttable}
        \caption{Patient demographics and number of patients in each stenosis category.}
        \rowcolors{2}{gray!25}{white}
        \footnotesize
        \begin{tabular}{l c c c c}
            \toprule
                \textbf{Groups} & \textbf{No Stenosis} (DS = 0\%) & \textbf{Moderate Stenosis} (0\%<DS<70\%)) & \textbf{Severe Stenosis} (DS$\geq$70\%) & \textbf{P-values} \\
            \midrule
                Patients & 20 & 12 & 7 &   \\
            Female & 12/20 (60\%) & 2/12 (17\%) & 1/7 (14\%) & 0.039 \\
            Male & 8/20 (40\%) & 10/12 (83\%) & 6/7 (86\%) & 0.039 \\
            Age & 55 ± 8 & 57 ± 10 & 59 ± 11 & 0.328 \\
            \bottomrule
        \end{tabular}

        \begin{tablenotes}
            \scriptsize
            \item DS = Diameter Stenosis, p-values were results of Chi-square tests for categorical variables or Welch’s t-tests for continuous variables. Values in the parentheses represent the percentage of the values in the corresponding category.
        \end{tablenotes}
        \label{tbl:demographics}
    \end{threeparttable}
}
\end{table}

\subsection{Computational Model and Boundary Conditions}

\begin{figure}
    \centering
    \includegraphics[width={.7\textwidth}]{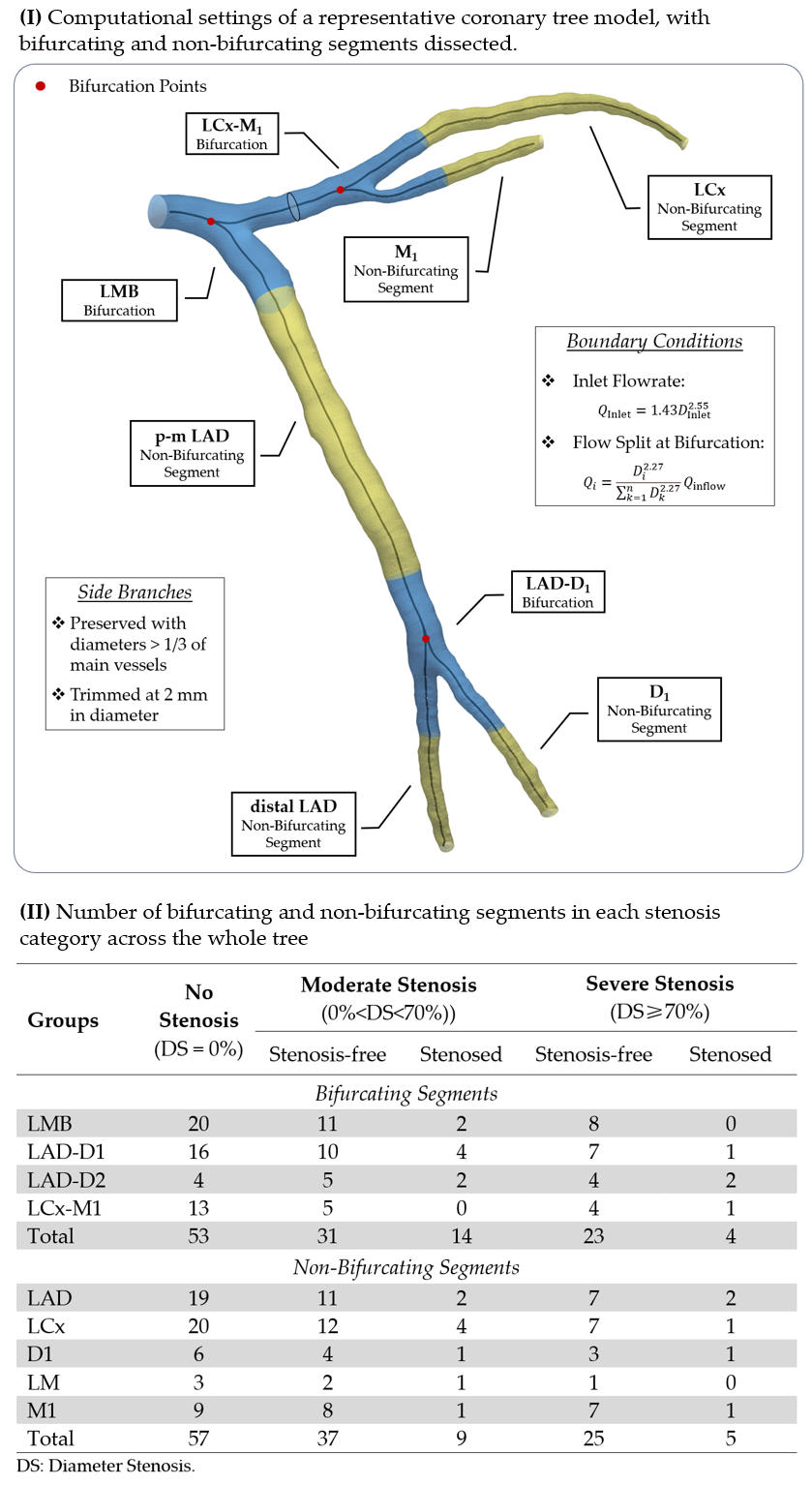}
    \caption{Patient demographics and breakdown of the whole left coronary tree into bifurcations (blue) and non-bifurcating (yellow) segments for sub-group analysis with each bifurcation defined as 10 mm along the vascular centreline proximal and distal to the bifurcation point. (LMB: Left Main Bifurcation, LAD: Left Anterior Descending artery, LCx: Left Circumflex artery, D1: the 1\textsuperscript{st} Diagonal artery, M1: the 1 \textsuperscript{st} Marginal artery, and p-m LAD indicates the proximal and middle segments of the LAD.)}
    \label{fig:breakdown}
\end{figure}

All distal coronary branches were trimmed if \textless2 mm in diameter due to the limited resolution of CTCA. Only side branches with diameters >1/3 of the main vessels were preserved. Bifurcating and non-bifurcating coronary tree segments were defined as 10 mm centreline length proximal and distal to the bifurcation point  \cite{Medrano-Gracia2016computational} (\textbf{Figure \ref{fig:breakdown}} [i]), resulting in 53 bifurcations and 57 non-bifurcating segments for the 20 no-stenosis cases, and 54 bifurcations (12 with stenoses) and 62 non-bifurcating segments (14 with stenoses) for the 19 moderately or severely stenosed cases (\textbf{Figure \ref{fig:breakdown}} [ii]). To examine the anatomical and haemodynamic differences by plaque severity, we used a Diameter Stenosis (DS) rate to classify non-stenosed, moderate and severe stenosis with DS=0\%, 0\%\textgreater DS\textless70\%, and DS$\geq$70\%, respectively. 
 
Since patient-specific coronary flow conditions were unavailable, we adopted a standard uniform velocity profile and waveform from literature \cite{Vlachopoulos2011McDonald's} as the inlet condition after scaling according to the patient-specific inlet diameter \textit{D}, deriving the scaled cycle-averaged flowrate \textit{Q} \cite{Giessen2011influence}:

\begin{equation}
    Q=1.43D^{2.55}.  
\end{equation}

The coronary haemodynamics were quantified assuming a resting state, with a flow split outflow strategy applied at each bifurcation, dividing the flowrate at the proximal main vessel into the daughter branches \cite{Giessen2011influence}:

\begin{equation}
    Q_i=\frac{D_i^{2.27}}{\sum_{k=1}^n D_k^{2.27} }Q_{\text{inflow}},  
\end{equation}

where $n$ is the number of daughter arteries at a bifurcation, and $i$ refers to the $i^{th}$ daughter branch, $D$ is the diameter averaged from the parameterised centreline points along its 10 mm length, as recommended as appropriate for atherosclerotic arteries where \textit{in vivo} data is unavailable \cite{schrauwen2016impact}, and outflow $Q_i$ is to be calculated. 

Computational meshes of each coronary model were generated using ICEM-CFD embedded in the ANSYS package (version 2023R1,  Canonsburg, USA), with the maximal sizes of the surface and volume elements determined as 0.1 and 0.2 mm, following a mesh sensitivity analysis. Five prismatic boundary layers adhering to the coronary wall were generated to resolve the near-wall blood flow. The coronary wall was assumed to be rigid, and a laminar fluid model was used with the maximal Reynolds number below 2,000.  The blood flow was modelled as incompressible, and the Carreau-Yasuda non-Newtonian fluid model \cite{Razavi2011Numerical} was used to reflect the shear-thinning behaviour of blood. 

\subsection{Coronary Anatomy and Haemodynamic Analysis}
An automated coronary shape analysis was performed on the 3D geometries with the centrelines calculated using an in-house code based on the Vascular Modelling ToolKit (VMTK, version 1.4 \cite{Antiga2012VMTK:}). For all non-bifurcating segments and bifurcation branches, \textit{i.e.} the Proximal Main Vessel (PMV), Distal Main Vessel (DMV), and Side Branch (SB), we quantified the average absolute curvature $\kappa_a$ to quantify vessel tortuosity, as recommended in recent literature \cite{Kashyap2022Accuracy}, following 
\begin{equation}
    \kappa_a = \frac{1}{L}\int_{s1}^{s2}\frac{\mid c^{'}(s)\times c^{''}(s)\mid} {\mid c^{'}(s)\mid ^3} ds,
\end{equation}
where $c(s)$ denotes the centreline parameterised along the coordinate $s$ of curve $c$, and $L$ represents the length of a curve considered. Furthermore, we calculated the mean Maximal Inscribed Sphere Diameter (MISD), and torsion $\tau_a$:
\begin{equation}
    \tau_a = \frac{1}{L}\int_{s1}^{s2}\frac{\mid (c^{'}(s) \times c^{''}(s))\cdot c^{'''}(s) }{\mid c^{'}(s) \times c^{''}(s)\mid ^2} ds,
\end{equation}
For the bifurcations, we additionally computed the inflow angle, bifurcation angle, and the Finet’s ratio \cite{Medrano-Gracia2016computational} calculated as
\begin{equation}
    \text{FR} = \frac{D_{\text{PMV}}}{D_{\text{DMV}} + D_{\text{SB}}}.
\end{equation}
A detailed definition of the considered anatomical measures is given in Table. \ref{tbl:definitions}. 
\begin{table}
\centering
\resizebox{\textwidth}{!}{
\begin{threeparttable}
    \caption{Definitions of coronary geometric parameters for bifurcating and non-bifurcating segments\textbf{.}}
    \rowcolors{2}{gray!25}{white}
     \begin{tabular}{p{3in}p{5in}}
        \toprule
           \textbf{Geometric Parameters } & \textbf{Definitions} \\
        \midrule
            \multicolumn{2}{c}{\textit{Non-Bifurcating Segments}}   \\   
            MISD& Mean Diameter of the largest spheres that fit inside the  coronary lumen at the points parameterised along the centreline\\
            Aboslute Curvature& 
                Mean $\kappa_a$ corresponding to the points parameterised along the centreline\\
            Torsion&  
Mean $\tau_a$ corresponding to the points  parameterised along the centreline\\
            \multicolumn{2}{c}{\textit{Bifurcating Segments}}   \\
            Inflow Angle& Angle with which a PMV enters the bifurcation plane, \textit{i.e.}, a least-square plane fitted to all the centreline points of the DMV and SB \\
            Bifurcation Angle& Angle of a bifurcation between the DMV and the SB \\
            PMV/DMV/SB Diameter& Mean diameter of the PMV, DMV, or SB\\
           
            PMV/DMV/SB Curvature&Mean aboslute curvature of the PMV, DMV, or SB\\
            
            PMV/DMV/SB Torsion&Mean torsion of the PMV, DMV, or SB\\
             
            Finet’s Ratio& 
Ratio of the mean PMV diameter to that of the mean DMV and SB
 diameters\\
        \bottomrule
        
     \end{tabular}

    \begin{tablenotes}
      \small
      \item  PMV: Parent main vessel, DVM: Daughter Main Vessel, SB: Side Branch  
    \end{tablenotes}
    \label{tbl:definitions}
\end{threeparttable}
}

\end{table}

Transient Computational Fluid Dynamics (CFD) simulations were performed using ANSYS-CFX for four cardiac cycles, with results taken from the fourth cycle to minimise transient start-up effects. A time step of 0.005 seconds was specified for the implicit 2\textsuperscript{nd} order temporal discretisation scheme. The criterion for convergence was set as 10\textsuperscript{-4} for the continuity and normalised velocity and pressure. We quantified the normalised luminal area exposed to adversely low TAESS (\%LowTAESS), high OSI (\%HighOSI), and RRT (\%HighRRT) due to their association with endothelial cell dysfunction and plaque development \cite{Peiffer2013Does}, calculated as:

\begin{equation}
    \text{TAESS} = \frac 1 T \int_0^T|\mathbf{\tau_w}|dt \\
\end{equation}
\begin{equation}
    \text{OSI} = \frac 1 2 \left(1- \frac{|\int_0^T \mathbf{\tau_w} dt|}{\int_0^T |\mathbf{\tau_w}| dt}\right)\\
\end{equation}
\begin{equation}
    \text{RRT} = \frac{1}{(1-2\cdot\text{OSI})\cdot\text{TAESS}}
\end{equation}

where $\tau_{w}$ is the flow-induced shear stress vector at the luminal wall, and \textit{T} denotes the cardiac cycle period. 

It is important to note that although low TAESS, high OSI and RRT are generally considered to have an adverse effect on the endothelial cells \cite{peiffer_does_2013}, different thresholds of each parameter have repeatedly been proposed in the literature with uncertainty around their patho-physiologically relevance. Thus, for comparisons between non-stenosed (DS = 0\%) and moderately or severely stenosed (DS\textgreater0\%) coronaries, we investigated and reported on all recommended thresholds for \%LowTAESS including 0.4, 0.5, 1.3, and 2.5 Pa denoted as, for example,  \%LowTAESS@0.5Pa \cite{Hartman2021definition,Chiastra2017Healthy, Xie2014Computation, Stone2018Role}. Similarly,  for \%HighOSI we studied 0.1 and 0.2 \cite{Chiastra2017Healthy,Xie2014Computation}. Since RRT is derived from TAESS and OSI, eight thresholds emerged for \%HighRRT, \textit{i.e.}, 0.50, 0.67, 0.96, 1.28, 2.50, 3.13, 3.33, and 4.17 Pa$^{-1}$. We found identical relationships regardless of the threshold chosen as presented in the results (and reported in detail in Appendix A) and thus have not expanded this consideration for the comparisons between moderately (0\% \textless DS \textless 70\%) and severely (DS $\geq$ 70\%) stenosed coronaries, where we instead studied only the most commonly used thresholds, \textit{i.e.}, \%LowTAESS@0.5Pa, \%HighOSI@0.1, and \%HighRRT@2.5Pa\textsuperscript{-1}. 

\subsection{Statistical Analysis}
The statistical analyses were conducted using the R language-based software JASP (version 0.17.3) \cite{JASPTeam2023JASP}. Continuous variables were expressed as mean and Standard Deviation (SD), and categorical variables were given as counts and percentages. Anatomic and haemodynamic differences between the non-, moderately, and severely stenosed arteries were studied for the whole coronary trees, and all bifurcation and non-bifurcating segments. We used a Shapiro-Wilk test to check for the normality of all distributions before using a Welch’s t-test for the normally distributed variables or a Mann-Whitney U-test for non-normally distributed variables. Both are considered suitable for the comparison of small samples. To account for the multiple comparisons considered here, we adjusted the p-values with a Bonferroni correction to reduce the chances of a false-positive result (type-I error) before interpreting their significance \cite{Dunn1961Multiple}, whereby, after the correction, a \textit{p} \textless 0.05* was considered statistically significant. A Receiver Operating Characteristics (ROC) curve was used to reveal the diagnostic performance if a parameter was significantly different between groups with a minimal sample size \textgreater 26, following a power estimation \cite{hanley_meaning_1982} . The number of patients considered allowed us to only measure the sensitivity, specificity, and area under the ROC curve (AUC) of a parameter in discriminating moderate or severe stenosed arteries (DS \textgreater 0\%) against non-stenosed arteries (DS = 0\%), but not moderately (0\% \textless DS>70\%) against severely stenosed arteries (DS$\geq$70\%). Correlations between the anatomical and haemodynamic metrics were examined using the Pearson correlation coefficient $r$.

\section{RESULTS}

\begin{figure}
    \centering
    \includegraphics[width={.7\textwidth}]{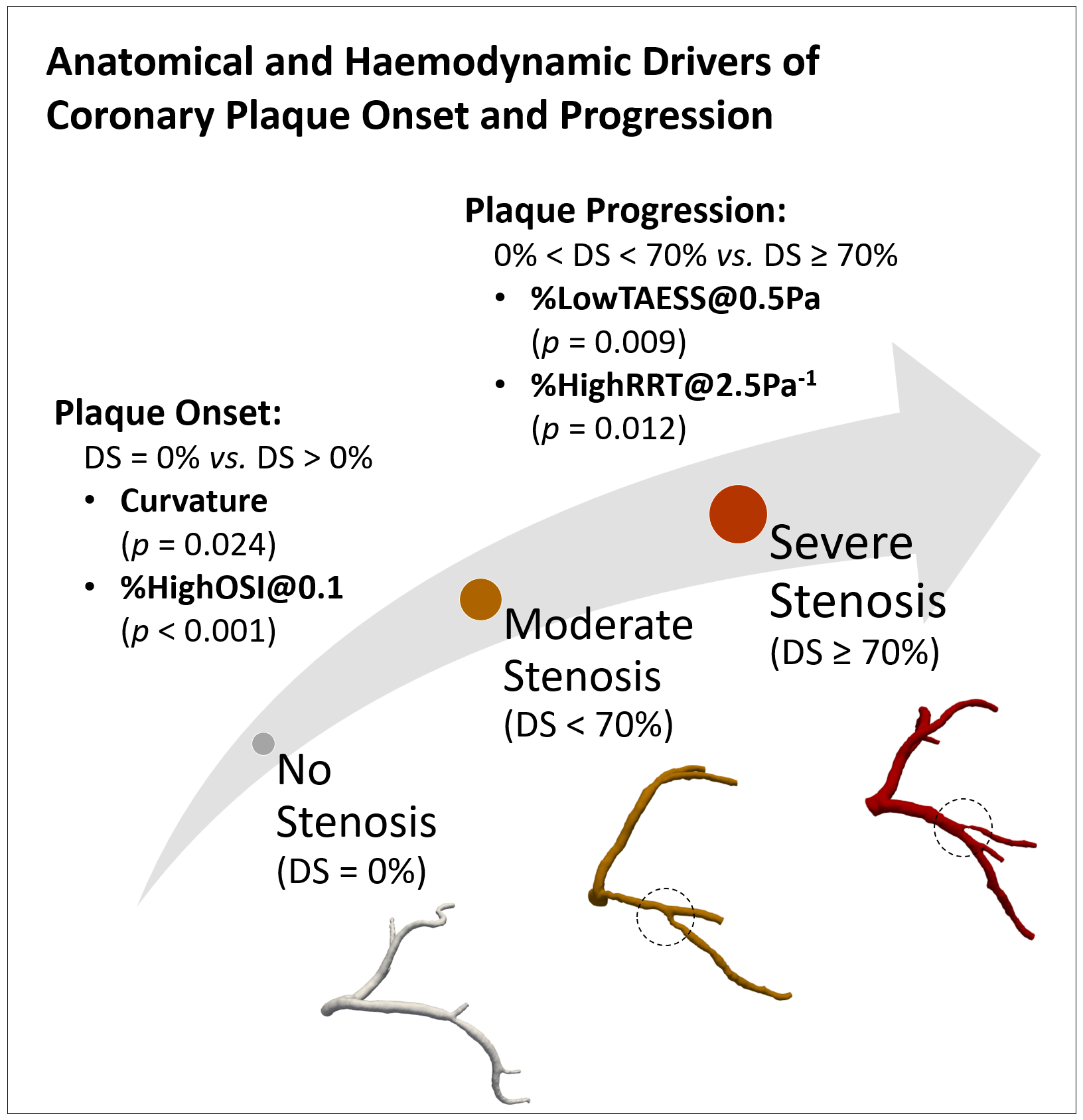}
    \caption{Comparisons suggest that the average absolute curvature and \%HighOSI statistically differed between non-stenosed and moderately or severely stenosed left coronary trees, revealing their link to plaque initiation, , whereas \%LowTAESS@0.5Pa and \%HighRRT@2.5\textsuperscript{-1} differed between arteries with moderate and severe stenoses, suggesting their role in plaque progression. DS: Diameter Stenosis.}
    \label{fig:illustration}
\end{figure}

We found significant differences with coronary trees with different stenosis rates in both coronary anatomy and haemodynamics (\textbf{Figure \ref{fig:illustration}}). Only the average absolute curvature and \%HighOSI (although with small absolute values) differed between arteries without and with stenosis when comparing the whole trees and non-bifurcating segments (DS = 0\% \textit{vs.} DS\textgreater 0\%), whilst the SB diameter and DMV torsion statistically differed between non-stenosed (DS = 0\%) and stenosed (DS\textgreater 0\%) bifurcations. The \%LowTAESS and \%HighRRT were the only parameters that statistically differed between moderately and severely stenosed arteries (0\% \textless DS \textless70\% \textit{vs.} DS $\geq$70\%). Haemodynamic thresholds did not appear to have an effect. Bifurcation-only studies may be misleading as other effects appear to take over with no statistically remarkable findings. 

All comparison results are listed in the Appendix, including all considerations of thresholds. Only key comparisons of statistical significance are reported below. 

\subsection{Plaque onset is affected by coronary curvature and high oscillatory shear stress}

\begin{figure}
    \centering
    \includegraphics[width={.9\textwidth}]{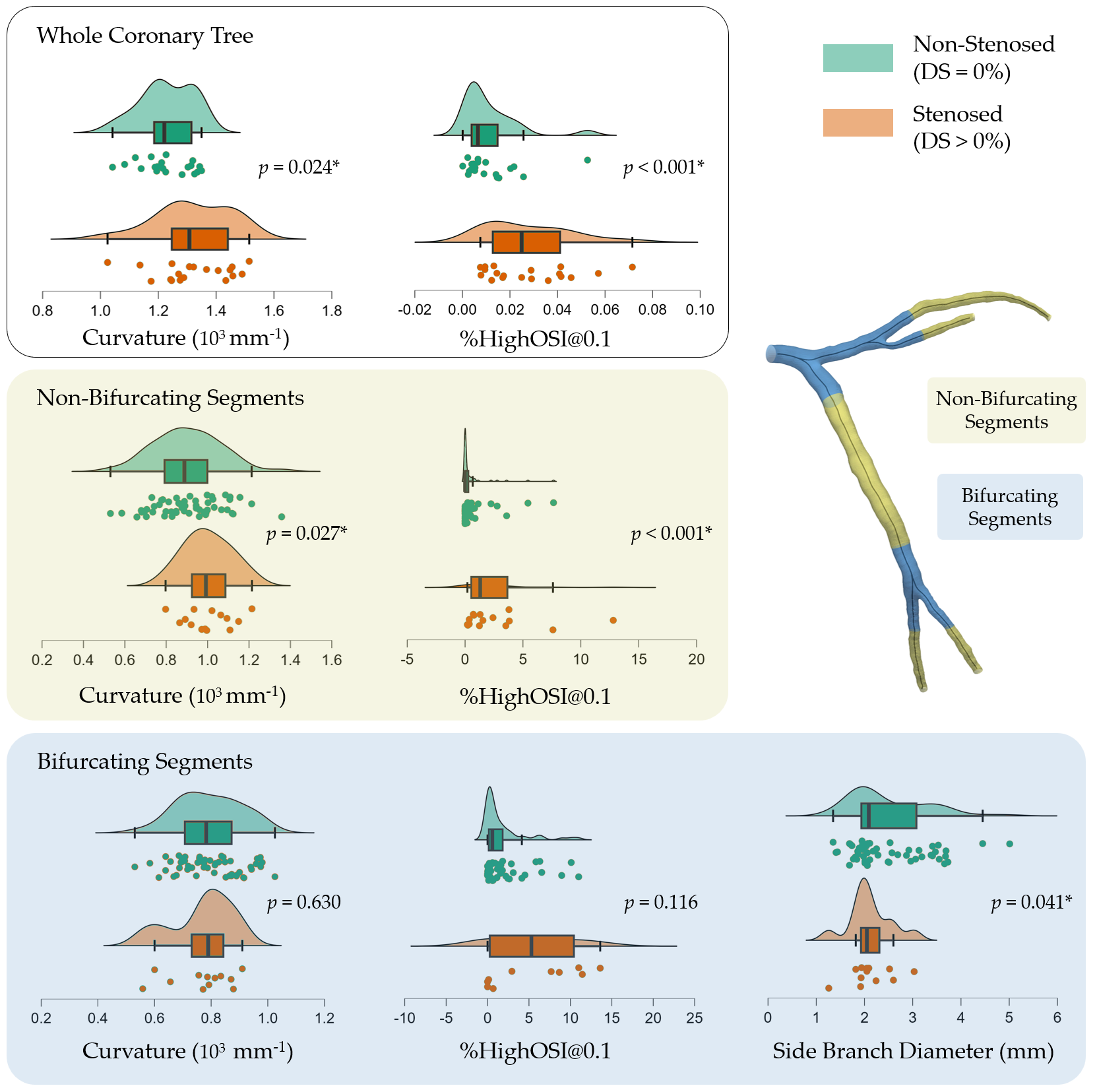}
    \caption{Comparisons of the whole coronary trees (top: n= 20 \textit{vs.} 19), non-bifurcating (middle: n=57 \textit{vs.} 14), and bifurcating segments (bottom: n=53 \textit{vs.} 12) for the differences in the average curvature (left) and \%HighOSI — here showing @0.1 threshold  for example (right) between patients with no stenosis (DS = 0\%) and with moderate or severe stenosis (DS \textgreater  0\%). Differences of statistical significance in the side-branch diameter are also shown for bifurcating segments (bottom-right) and results @0.2 threshold  for \%HighOSI are provided in the Appendix A. }
    \label{fig:onset}
\end{figure}

\begin{figure}
    \centering
    \includegraphics[width={.9\textwidth}]{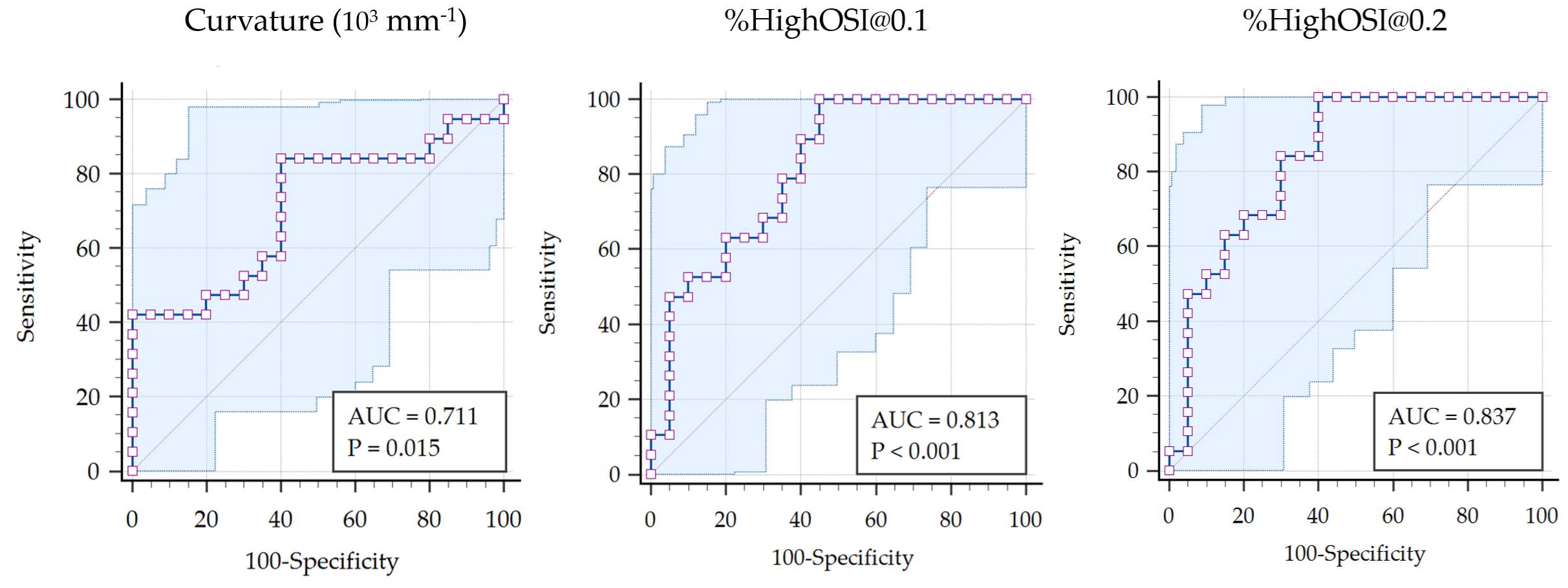}
    \caption{Diagnostic performance of coronary curvature, \%HighOSI@0.1 and \%HighOSI@0.2 in differentiating between non-stenosed and moderately or severely stenosed coronary arteries.}
    \label{fig:diagnostics}
\end{figure}

Absolute average curvature was the only anatomical characteristic exhibiting a statistically significant difference between the non-stenosed (DS = 0\%), and moderately or severely stenosed (DS \textgreater 0\%) cases when compared for both the whole coronary tree (DS = 0\%: 1.230 ± 0.090 \textit{vs.} DS \textgreater 0\%: 1.321 ± 0.131 m\textsuperscript{-1}, \textit{p} = 0.024), and the non-bifurcating segments (DS = 0\%: 0.902 ± 0.162 \textit{vs.} DS \textgreater 0\%: 1.001 ± 0.116 m\textsuperscript{-1}, \textit{p} = 0.027), as shown in \textbf{Figure \ref{fig:onset}}). Curvature had a good classification performance in identifying stenosed (DS \textgreater 0\%) coronary trees, with a sensitivity of 0.842 (95\% Confidence Interval [CI]: 0.604 to 0.966), a specificity of 0.600 (95\% CI: 0.361 to 0.809), and a AUC of 0.711 (95\% CI: 0.543 to 0.844), at the optimal cut-off of 1.227 m\textsuperscript{-1} (\textit{p} = 0.015, \textbf{Figure \ref{fig:diagnostics}}). 

However, this held not true for bifurcating segments within the same group classification, \textit{i.e.}, non-stenosed (DS=0\%) versus stenosed (DS \textgreater 0\%) bifurcations (\textit{p} \textgreater 0.057). This is most likely attributed to the complex interplay between coronary anatomy and haemodynamics around a bifurcation, which together affect the adverse ESS distribution. However, the SB diameter (\textit{p} = 0.041) and DMV torsion (\textit{p} = 0.024) were statistically different (\textbf{Appendix A}) , which warrants further analysis to confirm their role in a larger population. Within the anatomical features, coronary curvature negatively correlated with the diameter (\textit{r} = -0.474 and \textit{p} = 0.002), suggesting that smaller arteries tend to be more curved than larger ones. 

Similarly, \%HighOSI, regardless of the thresholds chosen (@0.1 or @0.2), was statistically different between the non-stenosed coronaries (DS = 0\%) and moderately or severely stenosed coronaries (DS \textgreater 70\%) , with the former having a lower \%HighOSI than the latter for the whole coronary tree (0.011 ± 0.012 \textit{vs. }0.028 ± 0.018, \textit{p} \textless 0.001) and non-bifurcating segments (0.821 ± 2.004 \textit{vs. }2.710 ± 3.438, \textit{p}\textless0.001, \textbf{Figure \ref{fig:onset}}).  However, OSI should be interpreted cautiously since the absolute values were generally small (\textless 0.04 and the luminal area affected by adverse OSI was \textless 15\%). In detecting stenosed (DS \textgreater 0\%) coronary trees, \%HighOSI@0.1 had a sensitivity and AUC of 0.999 (95\% CI: 0.824 to 0.999) and 0.813 (95\% CI: 0.656 to 0.920), and a specificity of 0.550 (95\% CI 0.315 to 0.769) at an optimal cut-off of 0.006 (\textit{p} \textless 0.001, \textbf{Figure \ref{fig:diagnostics}}). However, for bifurcations only, \%HighOSI, irrespective of the thresholds, manifested no statistical difference between non-stenosed and stenosed segments (p \textgreater0.116).

\subsection{Plaque progression is driven by RRT and TAESS}

\begin{figure}
    \centering
    \includegraphics[width={.7\textwidth}]{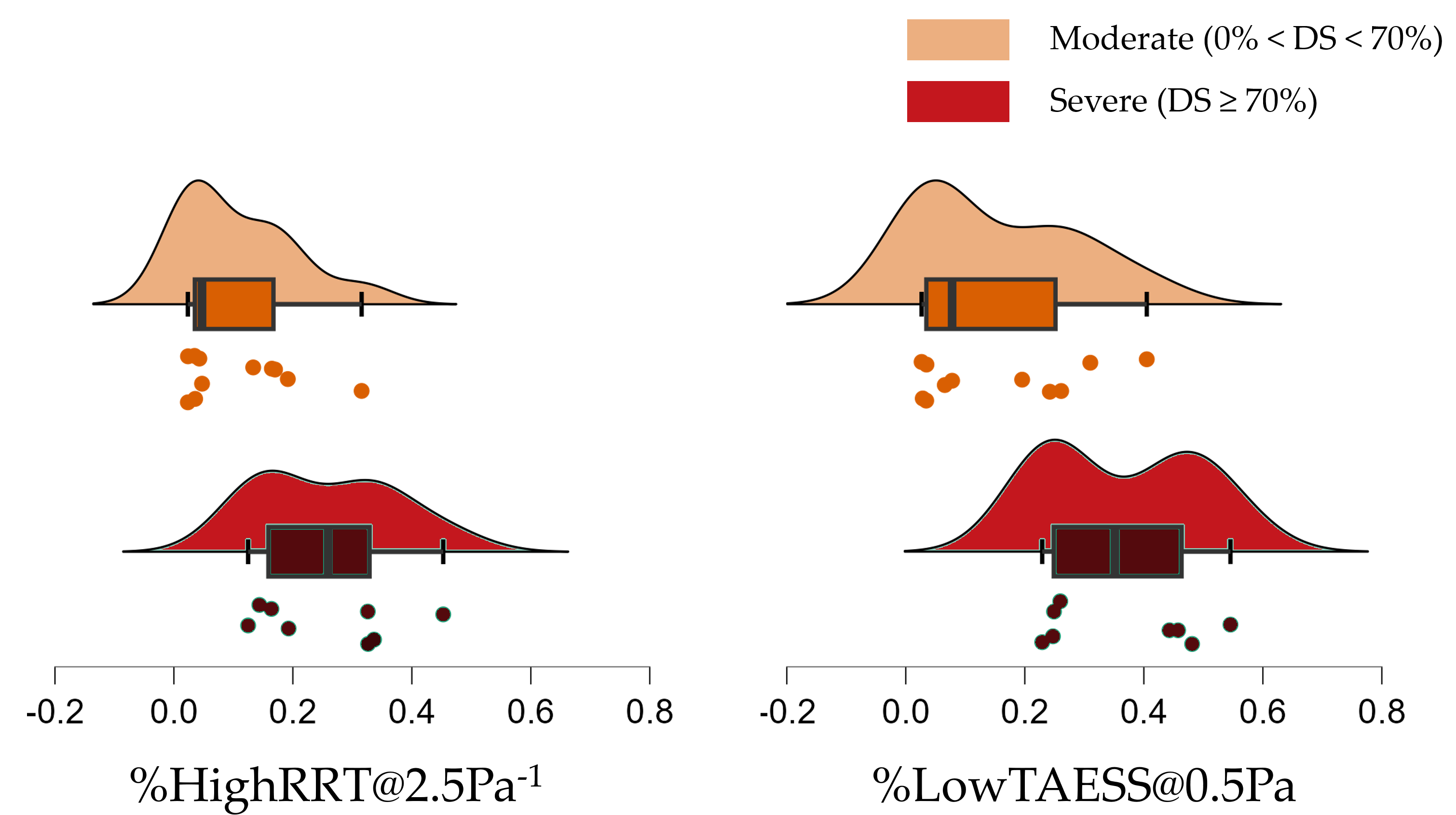}
    \caption{Differences in \%HighRRT@2.5Pa\textsuperscript{-1} and \%LowTAESS@2.5Pa between moderately (0\% < DS \textless 70\%) and severely (DS $\geq$ 70\%) stenosed coronaries. }
    \label{fig:progression}
\end{figure}

When comparing moderately (0 \textless DS \textless 70\%) to severely (DS \textgreater 70\%) stenosed cases only, we found a statistically larger \%HighTAESS@0.5Pa (moderate: 0.153 ± 0.135 \textit{vs.} severe: 0.364 ± 0.130 , \textit{p} = 0.009) and \%HighRRT@2.5Pa\textsuperscript{-1} (moderate: 0.108 ± 0.095 \textit{vs.} severe: 0.258 ± 0.118, \textit{p} = 0.012, \textbf{Figure \ref{fig:progression}}. \%HighOSI no longer significantly differed (e.g., @0.1 moderate: 0.026 ± 0.018, \textit{vs.} severe: 0.030 ± 0.020, \textit{p} = 0.665).  

Anatomical characteristics, including curvature, did not statistically differ across the coronary tree (\textit{p} \textgreater 0.272). Due to the limited sample size within the stenosed groups, \textit{i.e.}, only 4 severely stenosed bifurcations and 5 severely non-bifurcating segments total, a bifurcation-specific or non-bifurcating-segment-specific comparison was not performed.

\section{DISCUSSION}

We systematically compared differences in the arterial anatomy and blood flow between non-, moderately, and severely stenosed coronaries on the largest publicly available open-source dataset (ASOCA) of CAD patients. Our results revealed the potential shape and flow characteristics that may drive coronary plaque initiation (\textit{i.e.}, coronary curvature and potentially OSI) and progression (\textit{i.e.}, TAESS and RRT). The initial presentation of atherosclerotic plaques is coronary wall thickening towards the outside of the vascular lumen, extensive compensatory remodelling (also known as positive remodelling), which shows no lumen-protruding effects \cite{Samady2011Coronary}. Hence, early atherosclerosis is not easily detectable from measurement of the luminal dimensions through invasive coronary angiography. For this reason, key factors driving positive remodelling, including potentially coronary arterial shape and its local haemodynamics, are being extensively studied to gain a better understanding of plaque formation risks. Yet, findings are somewhat contradicting to date, with uncertainty around the exact mechanisms between local haemodynamics and anthogenesis.

For vascular anatomical characteristics, their link to plaque onset and progression remains inconclusive. Although various studies have suggested vascular tortuosity \cite{Liu2015Hemodynamic,Peng2016Impact}, its definition has not been consistent, and some metrics, \textit{e.g.}, the tortuosity index, are even incapable of capturing the actual bending of 3D vessels \cite{Kashyap2022Accuracy}. Clinical studies tend to measure tortuosity on 2D X-ray fluoroscopy images, defining tortuosity by the number of arterial bending with angles over a certain degree, \textit{e.g.}, 90° \cite{Han2022Association}, or classifying the shape of vessels concerning their 2D projections, \textit{e.g.}, C- or S-shaped \cite{Altintas2020Influence}. Even for tortuosity calculated by engineers on 3D centrelines, various equations have been used, including notably the tortuosity index \cite{Malvè2015Tortuosity} and average absolute curvature \cite{Hart1997Automated}, \textit{etc}. The inconsistencies in measuring tortuosity have thus disabled comprehensive cross-comparison of the findings from different studies. Moreover, tortuosity may also be affected by variations in the coronary diameter \cite{Shen2022Effect}, suggesting their potentially confounding effects on the local haemodynamics and, therefore, warrants further detailed analysis. Besides these centreline-based anatomical metrics, recent studies tend to test anatomical metrics based on the length or volume of the coronary arteries, \textit{e.g.}, the coronary artery volume index \cite{Benetos2020Coronary}, whereby a high prognostic efficacy for cardiovascular events was reported. 

Although TAESS and OSI have both been considered drivers of plaque onset \cite{Peiffer2013Does}, here we observed only OSI to be statistically differing between the non-stenosed and stenosed left coronary arteries. For blood flow investigations, OSI and atherosclerosis were first linked in post-mortem human carotid arteries \cite{Ku1985Pulsatile}, thereafter, the superior correlation of OSI to intimal thickness was confirmed and suggested for coronary arteries as well \cite{Hoi2011Correlation,xu_improvement_2020}. However, parallel studies disagreed, or at least found equivalent efficacy in TAESS and OSI — studies performed on patient-specific coronary models with plaque virtually removed suggested that low TAESS predict the most significant number of plaque locations \cite{Knight2010Choosing}. At the same time, OSI and RRT had fewer false negative predictions \cite{Rikhtegar2012Choosing}. Nonetheless, the merit of this approach is questionable since the arterial anatomy could have altered as a result of disease development, and virtual removal of plaques from stenosed arteries was reported with the potential of under- or overestimating the size of the healthy lumen \cite{Peiffer2013Does}. In another longitudinal study on human abdominal aortas, early atherosclerotic lesions were found to co-localise with both low TAESS and high OSI \cite{Buchanan2003Hemodynamics}. Yet, the reader is advised to exercise caution with interpreting such findings in the context of other vessels because flow regimes, especially for the aorta, are with much greater Reynolds numbers and thus may have distinct flow fields. 

Across these haemodynamic parameters, TAESS estimates the temporal average of the ESS magnitudes, whereby OSI characterises the degree of shear stress reversal and RRT identifies regions exposed to both low and oscillatory ESS. Thus, OSI and RRT may more comprehensively reflect the variation of both shear stress magnitude and direction on endothelial cells in a cardiac cycle. Still, this should be interpreted with caution since OSI is normally very low in coronaries, as demonstrated here, compared to that in carotid and aorta. In addition, there is a lack of fundamental studies on how OSI or RRT affect endothelial cells compared to ESS, whereby low TAESS, proximal or distal to a focal plaque, has been linked to plaque progression, and high TAESS at the plaque spot promote plaque erosion \cite{stone_prediction_2012}. Following similar protocols, our identification of the OSI difference before and after plaque onset, and the RRT difference between the moderate (0 \textless DS \textless 70\%) and severe (DS $\geq$ 70\%) stenoses, warrant further observations on large-scale cohorts, although the effects of RRT have just been endorsed by a recent animal experiment \cite{hoogendoorn_multidirectional_2020}. The insignificant differences for bifurcating segments as observed here also warrant further observation to examine the confounding effects from various anatomical and haemodynamic parameters on a larger patient groups. Bearing these uncertainties in mind, for future work, in addition to factors with established links to endothelial dysfunction, it may also aid to include flow characteristics that are known to prevent plaque formation. One such characteristic is the intensity of helical flow \cite{Morbiducci2015rational,Shen2021Secondary}, which reportedly has protective effects in animal studies \cite{de_nisco_impact_2020}. 

The absence of patient-specific boundary conditions for our flow simulations may be considered a limitation. However, coronary flow measurement is with great uncertainty and not routinely performed in cardiological practice, and thus may not be feasible for analysis and comparison of large retrospective registries. Non-invasive measurement of coronary artery flow relies on transthoracic Doppler echocardiography, whereas the depth and resolution of imaging, obstruction of the bones, \textit{etc}. have constrained its use within the arteries of LM, LAD and PDA. Invasive approaches capable of measuring the side-branch flows include Doppler sonography performed during Intravascular Ultrasound and TIMI Frame Count (TFC) that derives flow velocity from the counts of cine-angiographic frames \cite{Kunadian2009Use}. Here, we adopted a scaling method (\textit{n} = 2.55) to estimate the left coronary inflow based on the average diameter of the Left Main (LM) artery, which reportedly correlates well with IVUS-obtained flowrates (\textit{r}\textsuperscript{2} = 0.87) \cite{Giessen2011influence} and is the most common inflow strategy in this scenario. Thus, standardising the inflow assumption using the flow-diameter relation for population groups improves the simulation efficiency and assures physiologically realistic results.  

For outflow conditions, a lumped parameter model is usually applied to account for the stenosis-induced flow re-distribution within the epicardial arteries \cite{Taylor2013Computational}. This approach requires the total myocardial resistance to be derived from the volume of the heart muscle and is commonly used to capture the fractional flow reserve under the hyperaemic condition \cite{Kim2010Patient-Specific}. To ease the characterisation of flows for a large population, we previously compared the WSS and pressure drop between coronaries with and without stenosis, respectively under the resting and hyperaemic condition \cite{Zhang2022Comparison}. The results suggested a mild difference in the average WSS and pressure drop across stenosis of different severities, supporting a flow split method to be used at bifurcations under the resting condition.

This study has some limitations. Although we had used the largest open-source CAD dataset, the sample size is still relatively small due to its retrospective and single-centre nature, where the comparisons between small samples were accounted for by using the Welch's \textit{t}-tests rather than the Student \textit{t}-tests. Moreover, we have only included the most used haemodynamic metrics with established links to endothelial dysfunction. However, future work would benefit from including recently proposed novel parameters as well, \textit{e.g.}, the WSS topological skeleton. Finally, we have included only symptomatic patients with no, moderate, or severe stenoses at their first visit, without follow-ups to confirm their longitudinal changes. Virtual removal of plaques from stenosed arteries, as performed in prior studies\cite{Knight2010Choosing, Rikhtegar2012Choosing}, may enable a pseudo track of the local haemodynamics before plaque onset, whereas the uncertainties introduced in this manual procedure would be challenging to assess. However, we acknowledge the critical importance of longitudinal studies starting from the healthy state to plaque progression to understand better the link of vascular shape and blood flow to CAD trajectory.

\section{CONCLUSIONS}

Based on anatomic and blood flow analysis on the largest open-source dataset ASOCA, we found that the curvature and \%HighOSI statistically differed between coronaries with no stenosis (DS = 0\%) and with moderate or severe stenosis (DS \textgreater 0\%) (\textit{p} \textless 0.001), and TAESS and RRT between coronaries with moderate (0 \textless DS \textless 70\%) and severe (DS $\geq$ 70\%) stenoses (\textit{p} \textless 0.012). This suggests that curvature and OSI may predict plaque onset and  TAESS and RRT drive the progression of plaques from the moderate to severe stage. Our findings contribute to a clearer understanding of the anatomical and haemodynamic drivers of atherosclerotic plaque initiation and progression, and thus would have direct impact on the prevention, risk prediction and stratification of CAD.

\bibliographystyle{elsarticle-num}
\bibliography{refs}

\newpage
\appendix

\section{Anatomic and haemodynamic parameters compared at different thresholds}

We append four tables of differences in coronary anatomic characteristics (Table \ref{tbl:anatomics}) and \%LowTAESS (Table \ref{tbl:tawss}), \%HighOSI (Table \ref{tbl:osi}), and \%HIghRRT (Table \ref{tbl:rrt}) at various thresholds in literature between coronaries without (DS = 70\%) and with (DS \textgreater 70\%) stenoses.

\begin{table}
\centering
\resizebox{0.7\textwidth}{!}{
    \begin{threeparttable}
        \caption{Differences in coronary anatomic characteristics between non-stenosed (DS = 0\%) and moderately or severely stenosed (DS \textgreater 0\%) arteries.}
        \rowcolors{2}{gray!25}{white}
        \footnotesize
        \begin{tabular}{l l l l l l l}
            \toprule
                \textbf{Parameters} & \textbf{Groups} & \textbf{N} & \textbf{Mean} & \textbf{SD} & \textbf{SE} & \textbf{P-values} \\
            \midrule
                Curvature& Moderate or Severe Stenosis& 19 & 1.321 & 0.131 & 0.030 & \textbf{0.024*}\\
                 & No-Stenosis & 20 & 1.230 & 0.090 & 0.020 &  \\
                Torsion& Moderate or Severe Stenosis& 19 & 4.664 & 16.715 & 3.835 & 0.247\\
                 & No-Stenosis & 20 & -0.322 & 7.240 & 1.619 &  \\
                Diameter& Moderate or Severe Stenosis& 19 & 2.658& 0.528& 0.122& 0.084\\
                 & No-Stenosis & 20 & 2.896& 0.256& 0.058&  \\
            \bottomrule
        \end{tabular}

        \begin{tablenotes}
            \scriptsize
            \item $N$: number of samples; $SD$: standard deviation; $SE$: standard error; and P-values are results of Welch’s t-test or Mann-Whitney U-test depending on the distribution of samples.
        \end{tablenotes}
        \label{tbl:anatomics}
    \end{threeparttable}
}
\end{table}

\begin{table}
\centering
\resizebox{0.7\textwidth}{!}{
    \begin{threeparttable}
        \caption{Differences in \%LowTAESS at four common thresholds between non-stenosed (DS = 0\%) and moderately or severely stenosed (DS \textgreater 0\%) arteries.}
        \rowcolors{2}{gray!25}{white}
        \footnotesize
        \begin{tabular}{l l l l l l l}
            \toprule
                \textbf{Thresholds}& \textbf{Groups} & \textbf{N} & \textbf{Mean} & \textbf{SD} & \textbf{SE} & \textbf{P-values} \\
            \midrule
                \%LowTAESS@0.4Pa& Moderate or Severe Stenosis& 19 & 0.157 & 0.127 & 0.029 & 0.901 \\
                 & No-Stenosis & 20 & 0.174 & 0.168 & 0.038 &  \\
                \%LowTAESS@0.5Pa& Moderate or Severe Stenosis& 19 & 0.242 & 0.168 & 0.038 & 0.813 \\
                 & No-Stenosis & 20 & 0.273 & 0.184 & 0.041 &  \\
                \%LowTAESS@1.3Pa& Moderate or Severe Stenosis& 19 & 0.722 & 0.237 & 0.054 & 0.749 \\
                 & No-Stenosis & 20 & 0.799 & 0.118 & 0.026 &  \\
                \%LowTAESS@2.5Pa& Moderate or Severe Stenosis& 19 & 0.906 & 0.151 & 0.035 & 0.204 \\
                 & No-Stenosis & 20 & 0.961 & 0.042 & 0.009 &  \\
            \bottomrule
        \end{tabular}

        \begin{tablenotes}
            \scriptsize
            \item $N$: number of samples; $SD$: standard deviation; $SE$: standard error; and P-values are results of Welch’s t-test or Mann-Whitney U-test depending on the distribution of samples.
        \end{tablenotes}
        \label{tbl:tawss}
    \end{threeparttable}
}
\end{table}

\begin{table}
\centering
\resizebox{0.7\textwidth}{!}{
    \begin{threeparttable}
        \caption{Differences in \%HighOSI at two common thresholds between non-stenosed (DS = 0\%) and moderately or severely stenosed (DS \textgreater 0\%) arteries.}
        \rowcolors{2}{gray!25}{white}
        \footnotesize
        \begin{tabular}{l l l l l l l}
            \toprule
                \textbf{Thresholds} & \textbf{Groups} & \textbf{N} & \textbf{Mean} & \textbf{SD} & \textbf{SE} & \textbf{P-values} \\
            \midrule
                \%HighOSI@0.2& Moderate or Severe Stenosis& 19 & 0.013 & 0.011 & 0.003 & \textbf{<.001*}\\
                 & No-Stenosis & 20 & 0.004 & 0.006 & 0.001 &  \\
                \%HighOSI@0.1& Moderate or Severe Stenosis& 19 & 0.028 & 0.018 & 0.004 & \textbf{<.001*}\\
                 & No-Stenosis & 20 & 0.011 & 0.012 & 0.003 &  \\
            \bottomrule
        \end{tabular}

        \begin{tablenotes}
            \scriptsize
            \item $N$: number of samples; $SD$: standard deviation; $SE$: standard error; and P-values are results of Welch’s t-test or Mann-Whitney U-test depending on the distribution of samples.
        \end{tablenotes}
        \label{tbl:osi}
    \end{threeparttable}
}
\end{table}

\begin{table}
\centering
\resizebox{0.7\textwidth}{!}{
    \begin{threeparttable}
        \caption{Differences in \%HighRRT at eight common thresholds between non-stenosed (DS = 0\%) and moderately or severely stenosed (DS \textgreater 0\%) arteries.}
        \rowcolors{2}{gray!25}{white}
        \footnotesize

        \begin{tabular}{l l l l l l l}
            \toprule
                \textbf{Thresholds} & \textbf{Groups} & \textbf{N} & \textbf{Mean} & \textbf{SD} & \textbf{SE} & \textbf{P-values} \\
            \midrule
                \%HighRRT@0.5Pa\textsuperscript{-1}& Moderate or Severe Stenosis& 19 & 0.865 & 0.181 & 0.041 & 0.336\\
                 & No-Stenosis & 20 & 0.929 & 0.064 & 0.014 &  \\
                \%HighRRT@0.67Pa\textsuperscript{-1}& Moderate or Severe Stenosis& 19 & 0.78 & 0.216 & 0.049 & 0.667\\
                 & No-Stenosis & 20 & 0.855 & 0.099 & 0.022 &  \\
                \%HighRRT@0.96Pa\textsuperscript{-1}& Moderate or Severe Stenosis& 19 & 0.631 & 0.25 & 0.057 & 0.901\\
                 & No-Stenosis & 20 & 0.699 & 0.147 & 0.033 &  \\
                \%HighRRT@1.28Pa\textsuperscript{-1}& Moderate or Severe Stenosis& 19 & 0.482 & 0.239 & 0.055 & 0.857\\
                 & No-Stenosis & 20 & 0.537 & 0.181 & 0.041 &  \\
                \%HighRRT@2.5Pa\textsuperscript{-1}& Moderate or Severe Stenosis& 19 & 0.171 & 0.128 & 0.029 & 0.728\\
                 & No-Stenosis & 20 & 0.188 & 0.166 & 0.037 &  \\
                \%HighRRT@3.13Pa\textsuperscript{-1}& Moderate or Severe Stenosis& 19 & 0.107 & 0.088 & 0.02 & 0.807\\
                 & No-Stenosis & 20 & 0.117 & 0.144 & 0.032 &  \\
                \%HighRRT@3.33Pa\textsuperscript{-1}& Moderate or Severe Stenosis& 19 & 0.093 & 0.077 & 0.018 & 0.857\\
                 & No-Stenosis & 20 & 0.102 & 0.136 & 0.03 &  \\
                \%HighRRT@4.17Pa\textsuperscript{-1}& Moderate or Severe Stenosis& 19 & 0.056 & 0.048 & 0.011 & 0.428\\
                 & No-Stenosis & 20 & 0.06 & 0.095 & 0.021 &  \\
            \bottomrule
        \end{tabular}

        \begin{tablenotes}
            \scriptsize
            \item $N$: number of samples; $SD$: standard deviation; $SE$: standard error; and P-values are results of Welch’s t-test or Mann-Whitney U-test depending on the distribution of samples. 
        \end{tablenotes}
        \label{tbl:rrt}
    \end{threeparttable}
}
\end{table}

Results suggest that Curvature is the only geometrical parameter exhibiting statistically significant difference between non-stenosed (DS = 0\%) and moderately or severely stenosed (DS \textgreater 0\%) coronary arteries, and \%HighOSI consistently differed between the the two groups regardless of the thresholds chosen to determine adverse haemodynamic conditions. However \%LowTAESS and \%HighRRT consistently manifested no statistically significant differences.

\section{Comparisons of the dissected coronary bifurcations and non-bifurcating segments for differences in the anatomic and haemodynamic characteristics}

We append two tables for the differences in the bifurcating (Table \ref{tbl:per-bifur}) and non-bifurcating(Table \ref{tbl:per-straight}) segments in terms of the coronary anatomic characteristics and \% wall areas exposed to adverse haemodynamics at the most used thresholds, \textit{i.e.}, 0.5 Pa for low TAESS, 0.1 for OSI, and 2.5 Pa\textsuperscript{-1} for RRT.

\begin{table}
\centering
\resizebox{0.7\textwidth}{!}{
    \begin{threeparttable}
        \caption{Comparison of coronary shape and haemodynamics between non-stenosed (DS = 0\%) and moderately or severely stenosed (DS \textgreater 0\%) bifurcations.}
        \rowcolors{2}{gray!25}{white}
        \footnotesize

        \begin{tabular}{l l l l l l l}
            \toprule
                \textbf{Parameters} & \textbf{Groups} & \textbf{N} & \textbf{Mean} & \textbf{SD} & \textbf{SE} & \textbf{P-values} \\
            \midrule
                Anatomical Characteristics & & & & & &    \\
                IA& Moderate or Severe Stenosis& 12 & 161.583 & 8.273 & 2.388 & 0.058\\
                 & No-Stenosis & 53 & 155.786 & 12.297 & 1.643 &  \\
                BA& Moderate or Severe Stenosis& 12 & 57.083 & 28.125 & 8.119 & 0.742\\
                 & No-Stenosis & 53 & 59.964 & 20.657 & 2.760 &  \\
                PMV diameter& Moderate or Severe Stenosis& 12 & 3.380 & 0.796 & 0.230 & 0.312\\
                 & No-Stenosis & 53 & 3.638 & 0.654 & 0.087 &  \\
                DMV diameter& Moderate or Severe Stenosis& 12 & 2.942 & 0.586 & 0.169 & 0.424\\
                 & No-Stenosis & 53 & 3.103 & 0.754 & 0.101 &  \\
                SB diameter& Moderate or Severe Stenosis& 12 & 2.121 & 0.445 & 0.129 & \textbf{0.041*}\\
                 & No-Stenosis & 53 & 2.479 & 0.811 & 0.108 &  \\
                FR& Moderate or Severe Stenosis& 12 & 0.678 & 0.160 & 0.046 & 0.851\\
                 & No-Stenosis & 53 & 0.669 & 0.097 & 0.013 &  \\
                PMV curvature& Moderate or Severe Stenosis& 12 
& 0.677& 0.198& 0.028&0.630\\
                & No-Stenosis & 53 & 0.668& 0.264& 0.041&\\
                 DMV curvature& Moderate or Severe Stenosis& 12 
& 0.775& 0.174& 0.024&0.995\\
                 & No-Stenosis & 53 & 0.769& 0.222& 0.032&\\
                 SB curvature& Moderate or Severe Stenosis& 12 
& 0.914& 0.286& 0.039&0.057\\
                 & No-Stenosis & 53 & 0.980& 0.316& 0.047&\\
                 PMV torsion& Moderate or Severe Stenosis& 12 
& 3.073& 1.302& 0.190&0.065\\
                 & No-Stenosis & 53 & 2.674& 1.892& 0.299&\\
                 DMV torsion& Moderate or Severe Stenosis& 12 
& 3.428& 1.567& 0.215&\textbf{0.024*}\\
                 & No-Stenosis & 53 & 2.771& 1.234& 0.176&\\
                 SD torsion& Moderate or Severe Stenosis& 12 
& 3.207& 1.248& 0.173&0.057\\
                 & No-Stenosis & 53 & 2.760& 1.083& 0.161&\\
                Blood Flow Characteristics & & & & & &    \\
                \%LowTAESS@0.5Pa& Moderate or Severe Stenosis& 12 & 22.400 & 18.650 & 5.898 & 0.700\\
                 & 
No-Stenosis & 53 & 26.627 & 20.983 & 2.882 &  \\
                \%HighOSI@0.1& Moderate or Severe Stenosis& 12 & 5.614 & 5.426 & 1.716 & 0.116\\
                 & No-Stenosis & 53 & 1.714 & 2.630 & 0.361 &  \\
                \%HighRRT@2.5Pa\textsuperscript{-1})& Moderate or Severe Stenosis& 12 & 20.679 & 17.353 & 5.487 & 0.903\\
                 & No-Stenosis & 53 & 22.446 & 20.770 & 2.853 &  \\
            \bottomrule
        \end{tabular}

        \begin{tablenotes}
            \scriptsize
            \item $N$: number of samples; $SD$: standard deviation; $SE$: standard error; and P-values are results of Welch’s \textit{t}-test or Mann-Whitney \textit{U}-test depending on the distribution of samples. IA: inflow angle; BA: bifurcation angle; PMV: proximal main vessel diameter; DMV: distal main vessel diameter; SB: side-branch diameter; and FR: Finet’s ratio.
        \end{tablenotes}
        \label{tbl:per-bifur}
    \end{threeparttable}
}
\end{table}

\begin{table}
\centering
\resizebox{0.7\textwidth}{!}{
    \begin{threeparttable}
        \caption{Comparison of coronary shape and haemodynamics between moderate or severe (DS = 0\%) and stenosed (DS \textgreater 0\%) non-bifurcating segments.}
        \rowcolors{2}{gray!25}{white}
        \footnotesize

        \begin{tabular}{l l l l l l l}
            \toprule
               \textbf{Parameters}& \textbf{Groups}& \textbf{N}& \textbf{Mean}& \textbf{SD}& \textbf{SE}& \textbf{P-values}\\
            \midrule
                Anatomical Characteristics & & & & & &    \\
                Curvature& Moderate or Severe Stenosis& 14 & 1.001 & 0.116 & 0.031 & \textbf{0.027*}\\
                 & No-Stenosis & 57 & 0.902 & 0.162 & 0.021 &  \\
                Torsion& Moderate or Severe Stenosis& 14 & 3.095 & 0.681 & 0.182 & 0.072\\
                 & No-Stenosis & 57 & 3.509 & 0.717 & 0.095 &  \\
                Diameter& Moderate or Severe Stenosis& 14 & 2.594& 0.348& 0.148& 0.303\\
                 & No-Stenosis & 57 & 2.686& 0.284& 0.064&  \\
                Blood Flow Characteristics & & & & & &  \\
                \%LowTAESS@0.5Pa& Moderate or Severe Stenosis& 14 & 15.180 & 18.447 & 1.792 & 0.129\\
                 & No-Stenosis & 57 & 20.968 & 18.669 & 4.820 &  \\
                \%HighOSI@0.1& Moderate or Severe Stenosis& 14 & 0.821 & 2.004 & 0.195 & \textbf{<.001*}\\
                 & No-Stenosis & 57 & 2.710 & 3.438 & 0.888 &  \\
                \%HighRRT@2.5Pa\textsuperscript{-1})& Moderate or Severe Stenosis& 14 & 12.825 & 17.369 & 1.687 & 0.102 \\
                 & No-Stenosis & 57 & 18.103 & 17.750 & 4.583 &   \\
            \bottomrule
        \end{tabular}

        \begin{tablenotes}
            \scriptsize
            \item  $N$: number of samples; $SD$: standard deviation; $SE$: standard error; and P-values are results of Welch’s \textit{t}-test or Mann-Whitney \textit{U}-test depending on the distribution of samples. Tortuosities are the average absolute curvatures derived from the centrelines of each coronary tree.
        \end{tablenotes}
        \label{tbl:per-straight}
    \end{threeparttable}
}
\end{table}
Within the blood flow parameters, \%LowTAESS and \%HighRRT were generally  positively correlated across different thresholds (\textit{n} = 39, all \textit{p} \textless 0.002, with the maximal \textit{r} = 0.977), in contrast to the correlation between \%LowTAESS and \%HighOSI, with the strongest negative correlation observed between \%HighOSI@0.1 \textit{vs.} \%LowTAESS@2.5Pa (\textit{r} = -0.413, \textit{p} = 0.009). No other flow parameters showed significant differences when compared between the dissected bifurcating or non-bifurcating segments, irrespective of the thresholds chosen, including \%LowTAESS and \%HighRRT. 




\end{document}